\documentclass[a4paper,11pt]{article}

\usepackage{pos}
\usepackage[english]{babel} %per ridefinire convenzioni italiane
\usepackage[utf8]{inputenc}
\usepackage{amsthm} %per classe theorem etc
\usepackage{mathtools}
\usepackage{tensor} %per indici tensori
\usepackage{physics}
\usepackage{bm}
\usepackage{feynmp}
\usepackage{feynmp-auto}
\newcommand{\Cov}{\mathrm{Cov}}

\definecolor{myGreen}{rgb}{0.196, 0.603, 0.298}

\title{Inclusive semi-leptonic $B_{(s)}$ mesons decay at the physical $b$ quark mass}
\author*[a,b]{Alessandro Barone}
\author[d,e]{Shoji Hashimoto}
\author[a,b,c]{Andreas J\"uttner}
\author[d,e,f]{Takashi Kaneko}
\author[d,e]{Ryan Kellermann}

\affiliation[a]{School of Physics and Astronomy, University of Southampton, Southampton SO17 1BJ, UK}
\affiliation[b]{STAG Research Center, University of Southampton, Southampton SO17 1BJ, UK}
\affiliation[c]{CERN, Theoretical Physics Department, Geneva, Switzerland}
\affiliation[d]{High Energy Accelerator Research Organization (KEK), Ibaraki 305-0801, Japan}
\affiliation[e]{School of High Energy Accelerator Science, SOKENDAI (The Graduate University for Advanced Studies), Ibaraki 305-0801, Japan}
\affiliation[f]{Kobayashi-Maskawa Institute for the Origin of Particles and the Universe, Nagoya University, Aichi 464–8602, Japan}

\emailAdd{a.barone@soton.ac.uk}

\abstract{
We address the non-perturbative calculation of the decay rate of inclusive semi-leptonic $B_{(s)}$ mesons decays from lattice QCD. Precise theoretical Standard Model predictions are key ingredients in searches for new physics. This type of computation may eventually provide new insight into the long-standing tension between the inclusive and exclusive determinations of the CKM matrix elements $|V_{cb}|$ and $|V_{ub}|$.
We perform a pilot lattice computation for $B_s \rightarrow X_c \, l \nu_l$ and improve on existing techniques. The valence-quark masses in our simulations are approximately physical for the domain-wall strange and charm quarks as well as for the bottom quark, for which we use a relativistic heavy quark effective action. We report on our progress and discuss future plans towards a first study with fully controlled systematic effects.
}

\FullConference{%
The 39th International Symposium on Lattice Field Theory,\\
8th-13th August, 2022,\\
Rheinische Friedrich-Wilhelms-Universität Bonn, Bonn, Germany
}

\begin{document}
\maketitle
	
\section{Introduction}

At present, the $b$-quark sector of particle physics has shown several tensions with the Standard-Model predictions. One of these long-standing tensions involves the measured values of the CKM matrix elements $|V_{cb}|$ and $|V_{ub}|$ between exclusive and inclusive decays. 
For example, one of the most recent determination of $|V_{cb}|$ from the latest FLAG 21 review \cite{Aoki2021, Gambino2016} are
\begin{align*}
 |V_{cb}| &= (42.00\pm 0.64) \times 10^{-3} \quad \text{inclusive} \, ,\\
 |V_{cb}| &= (39.36\pm 0.68) \times 10^{-3} \quad \text{exclusive} \, .
\end{align*}
The exclusive determination comes from the decay $B\rightarrow D^{(*)}l\nu_l$, where the final-state hadron (either $D$ or $D^{*}$, assumed stable in QCD) is in its ground state. On the other hand, the inclusive determination is given from experimental measurements of all  possible semi-leptonic final states.
While there are many predictions on the exclusive channels from Lattice QCD, the inclusive sector is still rather unexplored, as such processes are difficult to treat on the lattice due to the presence of multiple hadrons in the final state.

In this work we perform a pilot study of the inclusive calculation of the semi-leptonic decays of $B_{s}$ mesons into charmed particles,
namely $B_{s}\rightarrow X_{c}l\nu_{l}$, following the work done in \cite{Hashimoto2017,Gambino2020}. In particular, we focus on developing a solid 
approach for the analysis, pushing forward and comparing two existing methods explained in the following sections.
A key aspect of our approach is the choice of the lattice action for the $b$ quark. In particular, we use the relativistic heavy quark (RHQ) action \cite{RHQFermilab,RHQColumbia1,RHQColumbia2}, which allows us to simulate the $b$ quark at its physical mass. The light, strange and charmed quarks are treated with a domain-wall fermion action and
 their masses are tuned to a value very close to the physical one. 

\section{Inclusive decays}

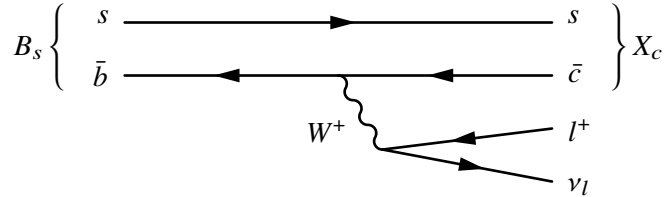
\begin{figure}[h!]
 \centering
 \begin{fmffile}{inclusive}
  \begin{fmfgraph*}(160, 60)
    \fmfset{arrow_len}{10}
    \fmfstraight
    \fmfleft{i4,i3,i2,i1}
    \fmfright{o4,o3,o2,o1}
    % fermions
    \fmffreeze
    \fmf{fermion}{i1,o1}
    \fmf{fermion}{o2,v2,i2}
    \fmffreeze
    \fmf{fermion,tension=1.5}{o3,v4,o4}
    \fmf{phantom,tension=1.8}{i4,v4}
    %\fmfv{l.d=20,l.a=180,l={$B_s$\mylbrace{32}{0}}}{B}
    %\fmfv{l.d=20,l.a=0,l={\myrbrace{32}{0}}$X_c$}{X}
    \fmflabel{$s$}{i1}
    \fmflabel{$\bar{b}$}{i2}
    \fmflabel{$\bar{c}$}{o2}
    \fmflabel{$s$}{o1}
    \fmflabel{$l^{+}$}{o3}
    \fmflabel{$\nu_l$}{o4}
    % boson
    \fmf{boson,label=$W^{+}$,label.side=left}{v4,v2}
    \fmfipair{B,X}
    \fmfiequ{B}{(-.2w,.47h)}
    \fmfiequ{X}{(1.2w,.47h)}
    \fmfiv{l=$B_s\hspace{1pt}\Bigg\{$, l.a=90}{B}
    \fmfiv{l=$\Bigg\}\hspace{1pt}X_c$, l.a=90}{X}
  \end{fmfgraph*}
 \end{fmffile}
 \caption{Feynman diagram for $B_s \rightarrow X_c\, l\nu_l$.}
 \label{fig:BtoXdiag}
\end{figure}

In this work we will focus on the decay $B_s \rightarrow X_c \, l\nu_{l}$ in Figure \ref{fig:BtoXdiag}.
The formalism is totally general and can be applied to other channels, such as $B \rightarrow X \, l\nu_{l}$, $D_{(s)}\rightarrow X \, l\nu_{l}$, etc.
The ground state for this decay is given by the $D_s$ meson, and $X_c$ stands for all the possible excited states. The expression for the differential decay rate is given by 
\begin{align*}
 \frac{\dd \Gamma}{\dd q^2 \dd q_0 \dd E_{l}} = \frac{G^2_F |V_{cb}|^2}{8\pi^3} L_{\mu\nu} W^{\mu\nu} \, ,
\end{align*}
where $W^{\mu\nu}$ is the hadronic tensor defined as
\begin{align}
 W^{\mu\nu} = \sum_{X_c}(2\pi)^3 \delta^{(4)}(p-q-r)\frac{1}{2E_{B_s}} 
 \bra{B_s(\bm{p})} J^{\mu\dagger}(-\bm{q})\ket{X_{c}(\bm{r})}  \bra{X_{c}(\bm{r})} J^{\nu}(\bm{q}) \ket{B_s(\bm{p})} 
 \label{eq:hadronicTensor}
\end{align}
and $L^{\mu\nu}$ is the leptonic tensor defined as
\begin{align}
 L^{\mu\nu} = p_{l}^{\mu}p_{\nu_l}^{\nu} +  p_{l}^{\nu}p_{\nu_l}^{\mu} - g^{\mu\nu} p_{l}\cdot p_{\nu_l} -i\epsilon^{\mu \alpha \nu \beta} p_{l\,,\alpha }p_{\nu_l\,,\beta} \, .
\end{align}
The hadronic tensor contains all the non-perturbative QCD effects, whereas the leptonic tensor is analytically known and contains kinematical factors, 
in particular the four-momenta of the lepton $p_l$ and the neutrino $p_{\nu_l}$, with $q=p_l + p_{\nu_l}$.
The sum $\sum_{X_c} \ket{X_{c}(\bm{r})}  \bra{X_{c}(\bm{r})}$ is over all hadronic intermediate states
integrated over all possible momenta $\bm{r}$ using the Lorentz invariant phase space integral.

From the above expression we can integrate over the electron energy $E_l$ to obtain the total decay rate
\begin{align}
 \Gamma &= \frac{G_F^2 |V_{cb}|^2}{24\pi^3} \int_{0}^{\bm{q}^2_{\rm max}} \dd \bm{q}^2\, \sqrt{\bm{q}^2} \, \bar{X}(\bm{q}^2)\, , \qquad
 \bar{X}(\bm{q}^2) \equiv \int_{\omega_{\rm min}}^{\omega_{\rm max}} \dd \omega \, k_{\mu\nu}W^{\mu\nu} \, ,
 \label{eq:total_Xbar}
\end{align}
where $\omega$ is the energy of the state $X_c$ and $k_{\mu\nu}=k_{\mu\nu}(\bm{q}, \omega)$ are kinematics factors 
originating from the leptonic tensor. The sum over the indices $\mu$, $\nu$ is implicit.
Considering that the $D_s$ meson is the lightest state of this process and imposing four-momentum conservation we get
$\bm{q}^{2}_{\rm max} = \left( \frac{M_{B_s}^2-M_{D_s}^2}{2M_{B_s}} \right)^2$, 
$\omega_{\rm min}=\sqrt{M_{D_s}^{2}+\bm{q}^2}$ and $\omega_{\rm max}=M_{B_s}-\sqrt{\bm{q}^2}$ for the limits of the integrals.

\section{Inclusive decays on the lattice}

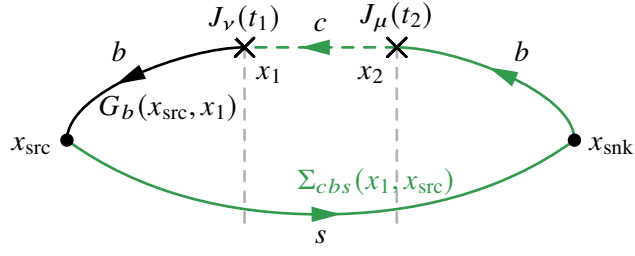
\begin{figure}[t!]
 \vspace{-0.7cm}
 \centering
 \begin{fmffile}{4pt}
  \begin{fmfgraph*}(190, 70)
     % box corners
    \fmfipair{tr,tc,tl,br,bc,bl}
    \fmfiequ{tl}{(0,h)}
    \fmfiequ{tc}{(.5w,h)}
    \fmfiequ{tr}{(w,h)}
    \fmfiequ{bl}{(0,-h)}
    \fmfiequ{bc}{(.5w,-h)}
    \fmfiequ{br}{(w,-h)} 
    % diagram
    \fmfipair{src,snk,t,tt,td,ttd,vm}
    \fmfiequ{src}{(0,0)}
    \fmfiequ{snk}{(w,0)}
    \fmfiequ{t}{(.35w,.5h)}
    \fmfiequ{tt}{(.65w,.5h)}
    \fmfiequ{td}{(.35w,-.5h)}
    \fmfiequ{ttd}{(.65w,-.5h)}
    \fmfiequ{vm}{(.5w,-.4h)}
    % labels
    \fmfipair{G,Gseq,xu,xd}
    \fmfiequ{G}{(.2w,.02h)}
    \fmfiequ{Gseq}{(.61w,-.38h)}
    \fmfiequ{xu}{(.40w,.22h)}
    \fmfiequ{xd}{(.60w,.22h)}
    % curved paths
    \fmfi{fermion, label=$b$}{t{left}  .. tension 1.5 .. {down}src}
    \fmfi{dashes_arrow, label=$c$, foreground=(0.196,, 0.603,, 0.298)}{tt .. t}
    \fmfi{fermion, label=$b$, foreground=(0.196,, 0.603,, 0.298)}{snk{up} .. tension 1.5 .. {left}tt}
    \fmfi{fermion, label=$s$, foreground=(0.196,, 0.603,, 0.298)}{src{bc-src} .. 1[src,vm] .. {snk-bc}snk}
    \fmfi{dashes, foreground=(0.7,,0.7,,0.7)}{t .. td}    
    \fmfi{dashes, foreground=(0.7,,0.7,,0.7)}{tt .. ttd}
    % point labels
    \fmfiv{d.sh=circle,d.f=1,d.siz=2thick,l=$x_{\rm src}$}{src}
    \fmfiv{d.sh=circle,d.f=1,d.siz=2thick,l=$x_{\rm snk}$}{snk}
    \fmfiv{d.sh=cross,d.f=1,d.siz=5thick, l=$J_{\mu}(t_2)$, l.a=90}{tt}
    \fmfiv{d.sh=cross,d.f=1,d.siz=5thick, l=$J_{\nu}(t_1)$, l.a=90}{t}
    \fmfiv{l=$G_b(x_{\rm src},,x_{1})$, l.a=90}{G}
    \fmfiv{l=\textcolor{myGreen}{$\Sigma_{cbs}(x_1,, x_{\rm src})$}, l.a=90}{Gseq}
    \fmfiv{l=$x_1$, l.a=90}{xu}
    \fmfiv{l=$x_2$, l.a=90}{xd}
  \end{fmfgraph*}
 \end{fmffile}
 \vspace{1.5 cm}
\caption{Diagram of the four-point correlator. Two propagators used for the contraction are depicted in the picture. The black one, $G_b(x_{\rm src}, x_1)$ is a propagator
for the $b$ quark from $x_1$ to $x_{\rm src}$. The green one, $\Sigma_{cbs}(x_1, x_{\rm src})$, is a sequential propagator that propagates the $s$ quark from $x_{\rm src}$ to $x_{\rm snk}$, 
the $b$ quark from $x_{\rm snk}$ to $x_2$ and the $c$ quark from $x_2$ to $x_1$.}
\label{fig:4pt}
\end{figure}

In order to address the computation of the hadronic tensor
\[
W_{\mu\nu} \sim \frac{1}{2E_{B_s}}\sum_{X_c}\bra{B_s}  J_{\mu}^{\dagger}\ket{X_c}\bra{X_c} J_{\nu}\ket{B_s} \, ,
\]
on the lattice we compute the four-point correlation function
\begin{align*}
 C_{\mu \nu}^{S J J S}\left(t_{\rm snk}, t_{2}, t_{1}, t_{\rm src}\right) =&
 \sum_{\bm{x}_{\rm snk}} e^{-i \bm{p}_{\rm snk}\cdot (\bm{x}_{\rm snk}-\bm{x}_{\rm src})} 
 \left\langle T 
 \left\{ \mathcal{O}_{B_s}^{S}\left(x_{\rm snk}\right) 
 \tilde{J}_{\mu}^{\dagger}\left(-\bm{q}, t_{2}\right) \tilde{J}_{\nu}\left(\bm{q}, t_{1}\right) 
   \mathcal{O}_{B_s}^{S \dagger}\left(x_{\rm src} \right) \right\} \right\rangle  \, ,
\end{align*}
where $\mathcal{O}_{B_s}$ is an interpolating operator for the the $B_s$ meson, and $\tilde{J}_{\nu}$ is the weak current in momentum space corresponding to the
$\bar{b}\rightarrow \bar{c}$ transition.
The superscripts ``$L$'' and ``$S$'' refer to ``local'' and ``smeared'' respectively. In particular, we smear the $B_s$ meson in order to improve the overlap with its
ground state. The diagram of the four-point function is depicted in Figure \ref{fig:4pt}.
In order to remove the dependence on the $B_s$ mass in the exponential we divide the four-point correlator by two $B_s$ two-point correlators as
\begin{equation}
 \frac{C^{SJJS}(t_{\rm snk},t_2 , t_1, t_{\rm src})}{C^{SL}(t_{\rm snk}, t_2)C^{LS}(t_1, t_{\rm src})} \, 
 \quad \stackrel{\substack{t_1-t_{\rm src} \gg 1 \\ t_{\rm snk}-t_2\gg 1}}{\longrightarrow} \quad
 \frac{\frac{1}{2M_{B_s}} \bra{B_s} \tilde{J}_\mu^{\dagger}(-\bm{q}, t_2) \tilde{J}_\nu(\bm{q}, t_1)  \ket{B_s} }
 {\frac{1}{2M_{B_s}}|\bra{0} \mathcal{O}_{B_s}^{L} \ket{B_s} |^2 } \, .
\label{eq:ratioC4/C2}
\end{equation}
Multiplying the above ratio by the amplitude factor and shifting $t=t_2-t_1$ we define the key quantity 
\begin{align*}
 C_{\mu\nu}(\bm{q}, t)  
  &\stackrel{t>0}{=} \sum_{\bm{x}} e^{i\bm{q}\cdot\bm{x}}\frac{1}{2M_{B_s}} \bra{B_s} J_{\mu}^{\dagger}(\bm{x},t) e^{-\hat{H}t}J_{\nu}(\bm{0},0) \ket{B_s} \\
  &= \frac{1}{2M_{B_s}} \bra{B_s} \tilde{J}_{\mu}^{\dagger}(-\bm{q},0)  e^{-\hat{H}t} \tilde{J}_\nu (\bm{q},0) \ket{B_s} \, .
\end{align*}
This is then the starting point to address the calculation of $\bar{X}(\bm{q}^2)$. Indeed, inserting $\delta (\hat{H} - \omega)$ and integrating over $\omega$ we get
\begin{align}
 \begin{split}
 C_{\mu\nu} (\bm{q}, t)
 &= \int_{0}^{\infty} \dd \omega \, \frac{1}{2M_{B_s}}
    \bra{B_s} \tilde{J}_{\mu}^{\dagger}(-\bm{q},0) \delta (\hat{H} - \omega) \tilde{J}_\nu (\bm{q},0) \ket{B_s} e^{-\omega t} \\
 &=  \int_{0}^{\infty} \dd \omega \, W_{\mu\nu}(\bm{q}, \omega) e^{-\omega t} \, .
 \end{split}
\end{align}
We immediately see from this representation that $W_{\mu\nu}$ is a sort of spectral function for the correlator $C_{\mu\nu}$, i.e.
\begin{align}
 W_{\mu\nu}(\bm{q}, \omega) \sim \sum_{X_c} \delta (\omega - E_{X_{c}}) \bra{B_s} J^{\dagger}_{\mu} \ket{X_c} \bra{X_c} J_{\nu} \ket{B_s} \, .
 \label{eq:WmunuSpectral}
\end{align} 
The determination of the hadronic tensor is then an ill-posed inverse problem, similar to the extraction of hadronic spectral densities from Euclidean
correlators.
However, we do not need to extract
the hadronic tensor directly, as we are interested in the quantity $\bar{X}(\bm{q}^2)$ in \eqref{eq:total_Xbar}. We try to address the calculation of the latter performing
some trivial but crucial steps:
\begin{align}
 \bar{X}(\bm{q}^2) = \int_{\omega_{\rm min}}^{\omega_{\rm max}} \dd \omega \, W^{\mu\nu}k_{\mu\nu} 
          = \int_{\omega_{0}}^{\infty} \dd \omega \, W^{\mu\nu}k_{\mu\nu} \theta(\omega_{\rm max} - \omega) 
          = \int_{\omega_{0}}^{\infty} \dd \omega \, W^{\mu\nu}K_{\mu\nu} \, ,
\end{align}
where $0 \leq \omega_0 \leq \omega_{\rm min}$ and $K_{\mu\nu}(\bm{q}, \omega) = k_{\mu\nu}(\bm{q}, \omega) \theta(\omega_{\rm max} - \omega)$ is 
the kernel operator. One has some freedom in choosing $\omega_0$ below the energy of the lowest energy state.
This will be useful later in the analysis. 

We now have to trade the unknown $\bar{X}(\bm{q}^2)$ with the lattice data $C_{\mu\nu}(t)$: this is possible as long as the kernel function
can be approximated
with polynomials in $e^{-a\omega}$ (we will set $a=1$ for simplicity) as
\begin{align*}
 K_{\mu\nu}(\bm{q}, \omega) \simeq c_{\mu\nu, 0} + c_{\mu\nu, 1}e^{-\omega} + \dots + c_{\mu\nu, N}e^{-N\omega} \, ,
\end{align*}
up to some maximum degree $N$. In this way we immediately see that
\begin{align}
 \begin{split}
 \bar{X}(\bm{q}^2) &\simeq c_{\mu\nu, 0} \int_{\omega_{0}}^{\infty} \dd \omega \, W^{\mu\nu} 
           + c_{\mu\nu, 1} \int_{\omega_{0}}^{\infty} \dd \omega \, W^{\mu\nu} e^{-\omega} + \dots 
           + c_{\mu\nu, N}\int_{\omega_{0}}^{\infty} \dd \omega \, W^{\mu\nu} e^{-N\omega} \\
         & = \sum_{j=0}^{N} c_{\mu\nu, j} C^{\mu\nu}(j) \, .
  \end{split}
\end{align}
It is important to notice that each term on the right hand side now directly corresponds to a correlator at a certain Euclidean time separation, and the final analysis will then
be limited by the available separations and the statistical noise of the data. 

We have now reduced the problem of computing the inclusive decay rate to the one of finding a suitable polynomial expression for the kernel
\[
   K_{\mu\nu}: [\omega_0, \infty) \; \rightarrow \; \mathbb{R} \, ,
\]
such that $K_{\mu\nu} \simeq \sum_{j}^{N} c_{\mu\nu,j} p_{j}(\omega)$, where $p_j(\omega)$ is a family of polynomials in $e^{-\omega}$. In order to 
work with a smooth function we trade the step function with a sigmoid
\[
   \theta_{\sigma}(x) = \frac{1}{1+ e^{-x/\sigma}} \, ,
\]
where $\sigma$ controls the sharpness of the step.

We now describe the two approaches followed in this work.

\subsection{Chebyshev polynomials}

The first approach is based on Chebyshev polynomials \cite{Weisse2006}, as they give the best approximation in term of the L$_\infty$-norm. The standard Chebyshev polynomials are defined in
\[
   T_{j}: \; [-1, 1]\; \rightarrow \; [-1,1] \quad \forall j \, .
\]
In this case, in order to match the domain of the target function, we consider a set of \textit{shifted Chebyshev polynomials in $e^{-\omega}$}, that
we indicate with $\tilde{T}_j(\omega)$. The kernel can now be approximated to some order $N$ as
\begin{align}
 K_{\mu\nu} = \sum_{j=0}^{N} c_{\mu\nu,j} \tilde{T}_j (\omega) \, , \qquad c_{\mu\nu, j}= \langle K_{\mu\nu}, \tilde{T}_j\rangle \, ,
\end{align}
where the coefficients are found projecting the target function into the Chebyshev polynomial basis, thanks to their orthogonality property. The inner product
$\langle \dots \rangle$ is defined to respect the orthogonality.

\subsection{Backus-Gilbert}
\label{sec:L2norm}

A second option is given by a variation of the Backus-Gilbert method \cite{Backus1968} as proposed in \cite{Hansen2019,Bulava2021}.
As long as the polynomial approximation
is concerned, the strategy of this method consists in minimising the L$_2$-norm 
\begin{align}
 A_{\mu\nu}[g] = \int_{\omega_0}^{\infty} \dd \omega \, \left[ K_{\mu\nu}(\bm{q}, \omega) - \sum_{j=1}^{N} g_{\mu\nu,j} e^{-j\omega} \right]^2
\end{align}
up to an order $N$.
The coefficients are then easily found by variational principle requiring $\dfrac{\delta A_{\mu\nu}}{\delta g_{\mu\nu,j}}=0$.

%%%%% PLOTS %%%%%

\begin{figure}[t!]
 \hbox{
 \hspace{-0.2 cm}
 \includegraphics[scale=0.31]{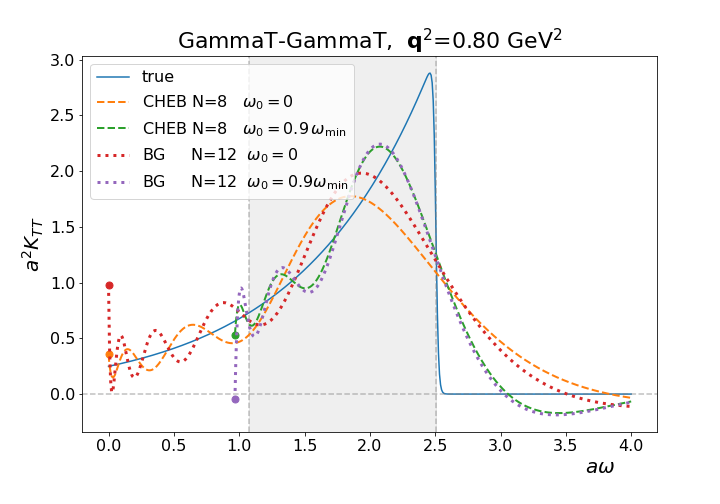}
 \hspace{-1 cm}
 \includegraphics[scale=0.31]{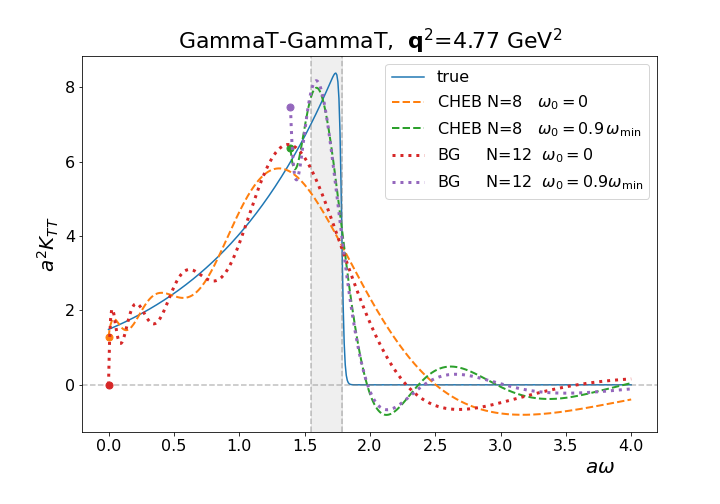}
 }
 \caption{Polynomial approximation of the kernel $K_{44}(\bm{q}, \omega)$, where ``4'' correspond to the time direction in Euclidean formulation.
 The plot on the left shows one of the smallest $\bm{q}^2=0.80 \, \text{GeV}^2$, whereas the plot on the right shows one of the largest momentum
 $\bm{q}^2=4.77 \, \text{GeV}^2$. The grey area corresponds to the kinematically allowed range $\omega_{\rm min}\leq \omega \leq \omega_{\rm max}$.}
 \label{fig:kernel_app}
 \vspace{-0.27 cm}
\end{figure}

The effect of the different approaches can be seen in Figure \ref{fig:kernel_app}.
The true kernel plotted with the blue solid line has a sigmoid with $\sigma=0.01$ and four different approximations are shown. In particular, there are two with Chebyshev and two with Backus-Gilbert, each of them
 starting the approximation either from $\omega_0=0$ or $\omega_0=0.9 \omega_{\rm min}$, as indicated also by the coloured dots in the plots. The different
choice for the order $N$ of the polynomial approximation, $N=8$ for Chebyshev and $N=12$ for Backus-Gilbert, is due to the limitations of the two strategies. In particular, the determination of the Chebyshev is strongly affected by the noise of the data and the fitting procedure, whereas the Backus-Gilbert is limited by the number of available time slices,
which depends on the choice of $t_2$ in the four-point correlation function.
The plot should give an idea of the final quality of the approximation for this specific kernel.
In particular, the one shown here is the one with the worst behaviour in terms of ``sharpness'' of the step.
A complete analysis should eventually take into account different choices of a smooth step function $\theta_{\sigma}$ and 
should make use of an extrapolation $\sigma \rightarrow 0$ in order to converge to the true kernel.

\section{Analysis strategies}
\label{sec:analysis_strat}

So far we have pointed out how to compute $\bar{X}(\bm{q}^2)$ from the lattice correlators. However, at this stage this is a naive approach: if we were to try to compute
the decay rate straight away the statistical errors would add up and lead to a very large error for the target quantity. We therefore need to devise some strategies to 
keep this under control in both approaches.

\subsection{Chebyshev polynomials}

For the Chebyshev polynomials, $\bar{X}(\bm{q}^2)$ would read
\begin{align}
 \begin{split}
 \bar{X}(\bm{q}^2) &\simeq c_{\mu\nu, 0} \int_{\omega_{0}}^{\infty} \dd \omega \, W^{\mu\nu} \tilde{T}_{0}(\omega) 
           + c_{\mu\nu, 1} \int_{\omega_{0}}^{\infty} \dd \omega \, W^{\mu\nu} \tilde{T}_{1}(\omega)  + \dots 
           + c_{\mu\nu, N}\int_{\omega_{0}}^{\infty} \dd \omega \, W^{\mu\nu} \tilde{T}_{N}(\omega) \, .
        % & = \sum_{j=0}^{N} c_{\mu\nu, j} \tilde{T}^{\mu\nu}_j 
  \end{split}
\end{align}
%
%where in the last step we defined $\tilde{T}^{\mu\nu}_j \equiv \int_{\omega_{0}}^{\infty} \dd \omega \, W^{\mu\nu} \tilde{T}_{j}(\omega)$. 
Using the polynomial representation of the Chebyshev, $\int_{\omega_{0}}^{\infty} \dd \omega \, W^{\mu\nu} \tilde{T}_{j}(\omega)$ can be written as a linear combination
of the correlator $C^{\mu\nu}(t)$ with $t=0,1,...,j$. Conversely, $C^{\mu\nu}(t)$ can be written as a linear combination of $\int_{\omega_{0}}^{\infty} \dd \omega \, W^{\mu\nu} \tilde{T}_{j}(\omega)$ with $j=0,1,...,t$. 

We can then exploit the fact that the Chebyshev polynomials are bounded \cite{Bailas2020a}, $| \tilde{T}_{j}(\omega)| \leq 1$, and redefine
\begin{align}
 \int_{\omega_{0}}^{\infty} \dd \omega \, W^{\mu\nu} \tilde{T}_{j}(\omega)
  \,  \rightarrow \,  \tilde{T}^{\mu\nu}_j \equiv
  \frac{\int_{\omega_{0}}^{\infty} \dd \omega \, W^{\mu\nu} \tilde{T}_{j}(\omega)}{\int_{\omega_{0}}^{\infty} \dd \omega \, W^{\mu\nu} \tilde{T}_{0}(\omega)}
  \, , \qquad
  |\tilde{T}^{\mu\nu}_j| \leq 1 \, .
\end{align}
We can now extract $\tilde{T}^{\mu\nu}_j$ from a \textit{fit with constraints} to the data, where $| \tilde{T}_{j}(\omega)| \leq 1$ is implemented
in terms of an augmented $\chi^2$.
After this, the final quantity can be calculated as
\begin{align}
 \bar{X}(\bm{q}^2) = \sum_{j=0}^{N} c_{\mu\nu, j} \,  \tilde{T}^{\mu\nu}_j \, \int_{\omega_{0}}^{\infty} \dd \omega \, W^{\mu\nu} \tilde{T}_{0}(\omega) \, .
\end{align}

\subsection{Backus-Gilbert}

The modified version of the Backus-Gilbert \cite{Hansen2019,Bulava2021} already includes a way to control statistical and systematics errors. In particular,
the idea is to consider an extra term on top of the one in Section \ref{sec:L2norm}, i.e. a functional $B_{\mu\nu}[g]$ defined as
\begin{align}
B_{\mu\nu}[g] = \sum_{i,j=1}^{N}g_{\mu\nu,i} \Cov[C_{\mu\nu}(i), C_{\mu\nu}(j)]  g_{\mu\nu,j} \, ,
\end{align}
which contains information about the data, hence the statistical error. The final functional to be considered is then
\begin{align}
 W_{\mu\nu,\lambda}[g] = (1-\lambda )\frac{A_{\mu\nu}[g]}{A_{\mu\nu}[0]} + \lambda \frac{B_{\mu\nu}[g]}{C_{\mu\nu}^2(0)} \, ,
\end{align}
where $\lambda$ is a parameter that controls the interplay between the systematic ($A_{\mu\nu}[g]$) and statistical ($B_{\mu\nu}[g]$) error.
The two terms are normalised in order to be of the same order of magnitude, such that their contribution can be weighted accordingly with the factor $\lambda$.

\section{Numerical setup}

We perform a pilot study using a $24^3\times 64$ lattice with 2+1-flavor domain-wall fermion (DWF)  Iwasaki gauge field ensembles
from RBC/UKQCD \cite{Allton2008} with lattice spacing $a\simeq 0.11 \,\text{fm}$ and pion mass $M_{\pi}\simeq 330 \,\text{MeV}$.
All the data have been generated with Grid \cite{Grid, GridProc} and Hadrons \cite{HadronsZenodo} software packages. Part of the fits in the analysis has been
performed using lsqfit \cite{lsqfit}.

The numerical parameters we use have been taken from the corresponding exclusive project from RBC/UKQCD \cite{Flynn2018,Flynn2019,Flynn2021}.
The strange quark is simulated using DWF \cite{Shamir1993,Furman1994}, whereas the charm quark is simulated by using the M\"obius DWF action \cite{Cho2015,Brower2017}.
Their masses are tuned such that the final hadrons have masses close to the physical ones.
The bottom quark has been
simulated using the Columbia formulation of the relativistic heavy quark (RHQ) action \cite{RHQColumbia1,RHQColumbia2},
which is based on the Fermilab heavy quark action \cite{RHQFermilab}. In particular, this
formulation allows to reduce the $b$ quark discretisation effects of order $\mathcal{O}((m_0a)^n)$, $\mathcal{O}((\bm{p}a)$ and $\mathcal{O}((\bm{p}a)(m_0a)^n)$
by tuning three non-perturbative parameters, one of this being the bare mass $m_0$.

For the computation we average over 60 different gauge configurations and the measurements are performed on 4 different source planes linearly spaced.
We use $\mathbb{Z}_2$ sources to improve the signal.
We induce 8 different momenta in the four-point functions 
using twisted boundary conditions \cite{DeDivitiis,Sachrajda2004} with same momentum in all the three directions. Considering 
$\bm{q} = \frac{2\pi}{L} \,\bm{n}$ in lattice units we have
$\bm{n}=(n_x, n_y, n_z) \equiv (\theta, \theta, \theta)$, where $\theta$ indicates the twist. We choose them such that all the momenta are linearly spaced in $\bm{q}^2$ and in particular $\theta_k= 1.90 \, \sqrt{\frac{k}{3}}$, where the factor $1.90$ is determined by the value of $\bm{q}^2_{\rm max}$ on 
the given ensemble.

\vspace{-0.15cm}
\section{Results}

\begin{figure}[t]
 \hbox{
 \includegraphics[scale=0.31]{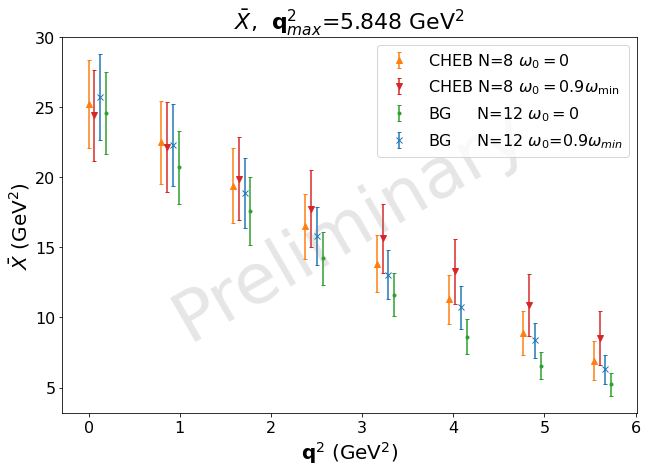}
 \includegraphics[scale=0.31]{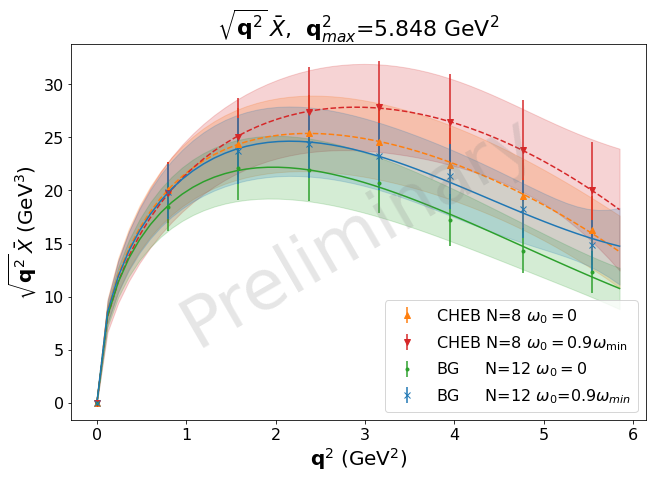}
 }
 \caption{The left plot shows the estimate of $\bar{X}(\bm{q}^2)$ with the two different strategies for 8 different $\bm{q}^2$.
 The right plot shows the same data on the left multiplied by $\sqrt{\bm{q}^2}$. The curves with error bands are determined through a
 a polynomial fit to interpolate the points and the area below the curves is directly proportional to the $\Gamma / |V_{cb}|^2$. }
 \label{fig:results}
\end{figure}

We now summarise the main results of this pilot study. In Figure \ref{fig:results} we show the result of $\bar{X}(\bm{q}^2)$ at different $\bm{q}^2$. The four data points
on each $\bm{q}^2$
correspond, as before, to the possible combinations of Chebyshev and Backus-Gilbert with either $\omega_0=0$ or $\omega_0=0.9 \omega_{\rm min}$. 
We can immediately see that all the points agree with each other in the full range. However, at higher $\bm{q}^2$ we note larger deviations; this is likely
due to the greater difference in the approximation as the kinematic range in $\omega$ shrinks, as shown in Figure \ref{fig:kernel_app}. 

The plot on the right shows $\sqrt{\bm{q}^2}\,\bar{X}(\bm{q}^2)$ and the area below the curves is directly proportional to the decay rate $\Gamma / |V_{cb}|^2$.
Since this is a purely qualitative study, we do not quote any number for the final decay rate. However, the value we obtain is in the right ballpark if compared with the $B$ meson decay rate assuming $SU(3)$ flavour symmetry. The final statistical error, at this preliminary stage,
is of order $10\%$ and all four values are in agreement within one standard deviation.

\section{Summary and outlook}

In this work we have shown that there are promising prospects for the study of inclusive decays on the lattice. At this stage, our study is qualitative
and focuses on setting the basis for a solid analysis to tackle the inverse problem arising in such processes. In particular, we compared
two different approaches, i.e. Chebyshev polynomials and Backus-Gilbert, and showed that their results are compatible within errors. 
A key aspect of our setup is also the choice of the $b$ quark, which is simulated at its physical mass through the RHQ action. This overcomes some of the issues that appear in \cite{Gambino2020, Gambino2022} and allows us to get a better idea on the outcome, since the numerical results should be in the right ballpark with respect the experimental data currently available.

Further improvements can be made in the analysis by developing new strategies to address
ill-posed inverse problems, starting from e.g. \cite{Bailas2020a,Bulava2021}.
The systematics associated with the polynomial approximation also need to be taken into account properly. Moreover, 
finite volume effects have been ignored so far and a careful check to justify this assumption
is required.

Once everything is in place it will be possible to plan for a full study on both $B_s$ and $B$, using multiple ensembles and taking into account
 the discretisation errors coming from the lattice, finite volume effects and continuum limit.

\section*{Acknowledgments}

This work used the DiRAC Extreme Scaling service at the University of Edinburgh, operated by the Edinburgh Parallel Computing Centre on behalf of the STFC DiRAC HPC Facility (www.dirac.ac.uk). This equipment was funded by BEIS capital funding via STFC capital grant ST/R00238X/1 and STFC DiRAC Operations grant ST/R001006/1. DiRAC is part of the National e-Infrastructure.
A.B. is supported by the Mayflower scholarship in the School of Physics and Astronomy of the University of Southampton.
The work of S.H. and T.K. is supported in part by JSPS KAKENHI Grant Number
22H00138 and 21H01085 respectively and by the Post-K and Fugaku supercomputer project through the
Joint Institute for Computational Fundamental Science (JICFuS).

%bibliography

\providecommand{\href}[2]{#2}\begingroup\raggedright\endgroup

\end{document}